\documentclass{jaa}
\usepackage{natbib}
\usepackage{psfrag,color,epsfig}
\usepackage{graphicx,graphics}	
\usepackage{url}
\usepackage{bm}
\usepackage{comment}
\usepackage{tabularx}
\usepackage{amsmath}
\usepackage{amssymb}
\usepackage{hyperref}

\usepackage{pdflscape}	% Landscape pages
\usepackage{multirow}
\usepackage{tabularx}  
\usepackage{physics}
\usepackage{xspace}

\usepackage[T1]{fontenc}
\usepackage{ae,aecompl}
\usepackage{newtxmath}

\def\HI{H\,{\scriptsize I}~}
\def\Hi{H\,{\scriptsize I}}

\def\xHI{x_{\rm H\,{\scriptsize I}}}

\def\bxHi{\bar{x}_{\rm HI}}
\def\Tb{{T_{\rm b}}}
\def\tTb{\tilde{T}_{\rm b}}

\def\x{{\bm{x}}}

\def\thetavec{{\bm{\theta}}}
\def\cl{{\mathcal C}_{\ell}}
\def\dl{{\mathcal D}_{\ell}}

\def\n{\hat{\bm{n}}}

\def\ncs{N_{\rm CS}}
\def\nrs{N_{\rm RS}}
\begin{document}\sloppy

\title{Quantifying and mitigating the effect of snapshot interval in light-cone Epoch of Reionization 21-cm simulations}

%%author names are separated by comma (,)
%%use \and before the last author name
%%use a * along with the number separated by comma
%% for the  author for correspondence
%%\textsuperscript{number} is used for affiliation
%%\affilOne, \affilTwo etc., upto \affilTwentyfive is possible
%%Please note the first letter after \affil is capitalised in the command
%%

\author{Suman Pramanick\textsuperscript{1,*}, Rajesh Mondal\textsuperscript{2} and Somnath Bharadwaj\textsuperscript{1}}
\affilOne{\textsuperscript{1}Department of Physics, Indian Institute of Technology Kharagpur, Kharagpur, West Bengal 721302, India\\}
\affilTwo{\textsuperscript{2}Department of Physics, National Institute of Technology Calicut, Calicut 673601, Kerala, India}

%%escape two column mode for title, affiliation and abstract
%%by giving \twocolumn command as shown

\twocolumn[{

\maketitle

%%include \corres to print the corresponding author Email id
\corres{suman21eor@gmail.com}

%%include \msinfo for
%%manuscript information such as
%%received, revised and accepted dates
%%
% \msinfo{1 January 2015}{1 January 2015}

%%abstract
\begin{abstract}
The Epoch of Reionization\,(EoR) neutral Hydrogen (\Hi)  21-cm signal evolves significantly along the line-of-sight\,(LoS)  due to the light-cone\,(LC) effect. It is important to accurately incorporate this in simulations  in order to correctly interpret the signal. The 21-cm LC simulations are typically produced by stitching together slices from a finite number $(\nrs)$ of ``reionization snapshot'', each corresponding to a different stage of reionization. In this paper, we have quantified the errors in the 21-cm LC simulation due to the finite value of $\nrs$. We show that this can introduce large discontinuities $(> 200 \%)$ at the stitching boundaries when $\nrs$ is small $(=2,4)$ and the mean neutral fraction jumps by $\delta \bxHi =0.2,0.1$ respectively at the stitching boundaries. This drops to $17 \%$ for  $\nrs=13$ where $\delta \bxHi=0.02$. We find that we can achieve $\delta \bxHi \le 0.01$ with  $\nrs =26$, and we use this as the reference for comparing the other simulations. 
We present and also validate a method for mitigating this error by increasing $\nrs$ without a proportional increase in the computational costs which are mainly incurred in generating the dark matter and halo density fields. Our method generates these fields only at a few redshifts, and interpolates them to generate reionization snapshots at closely spaced redshifts. We use this to generate 21-cm LC simulations with $\nrs=51,101$ and $201$, and show that the errors go down  as $\nrs^{-1}$.
\end{abstract}

%%insert keywords separated by 3 hyphens using \keywords{words}

}]
%%close the twocolumn escape here

%%include \doinum{number}for the DOI number in the header
%%include \volnum{number} for the volume number in the header
%%include \year{yyyy} for  year of publication in the header
%%include \pgrange{num--num} page range of article in the header
%%include \artcitid{num} for the article citation id
%%include \lp to print last page of the article
%%include \setcounter{page}{pagenum} for the exact starting page of the article

\doinum{12.3456/s78910-011-012-3}
\artcitid{\#\#\#\#}
\volnum{000}
\year{0000}
\pgrange{1--}
\setcounter{page}{1}
\lp{1}

\section{Introduction}\label{sec:intro}

The Epoch of Reionization (EoR) is the period in the history of the universe when the diffuse hydrogen in the Intergalactic Medium (IGM) underwent a transition from neutral to ionized. It is one of the least understood phases in the evolution history of our Universe. Our current understanding of the EoR is largely based on a few indirect observations. Observations of quasar absorption spectra indicate that the IGM is nearly completely ionized at low redshifts ($z < 5.5$)  with a mean neutral  hydrogen (H~{\scriptsize I}) fraction of $\bxHi \sim 10^{-5} - 10^{-4}$ \citep{becker2001evidence, fan2002evolution, bosman2018new, bosman2021comparison, eilers2018opacity, yang2020measurements,Zhu_2022}. These observations also find a rapid increase in the  Ly-$\alpha$ and Ly-$\beta$  effective optical depth at $z \ge 5.5$. This indicates a nearly completely ionized IGM by $z \sim 5.5$ \citep{mcgreer2014model}. Observations of the redshift evolution of the clustering and the luminosity function of Ly$\alpha$ emitters (LAEs) also suggest a rapid increase in the neutral fraction at $z > 5.5$ \citep{ouchi2010statistics, faisst2014spectroscopic, konno2014accelerated,ota2017new}. The analysis of Cosmic Microwave Background (CMB) observations provide estimates of the Thomson scattering optical depth $\tau_{\rm Th}$, which imposes constraints on models of reionization. The latest measurement by \cite{collaboration2020planck} indicates that reionization started around $z\sim 12$. These indirect observations together constrain reionization to the redshift range $6 \leq z \leq 12$ \citep{robertson2013new, robertson2015cosmic, mondal2018, mitra2017cosmic,mitra2018first, dai2019constraining}, with the possibility that it may extend as late as  $z \sim 5.5$.   However, the exact timing and duration of the EoR are still not well understood.

The redshifted 21-cm signal arising from the hyperfine transition in the ground state of \HI is the most promising direct probe of the EoR (see e.g. \citep{FURLANETTO2006181, pritchard201221}). A considerable amount of effort is currently underway to detect this signal by either measuring the mean sky-averaged signal (i.e. global signal) \citep{bowman2008toward, bowman2018absorption, singh2018saras} or  measuring the angular and spectral fluctuations using radio interferometers, such as the Giant Metrewave Radio Telescope (GMRT)\footnote{\url{http://www.gmrt.ncra.tifr.res.in}} \citep{swarup1991, ghosh2012characterizing, Paciga2013}, Low Frequency Array (LOFAR)\footnote{\url{http://www.lofar.org}} \citep{van2013lofar, yatawatta2013initial}, Precision Array for Probing the Epoch of Reionization (PAPER)\footnote{\url{http://eor.berkeley.edu}} \citep{parsons2014new, ali201564, jacobs2015multiredshift}, Murchison Widefield Array (MWA)\footnote{\url{http://www.haystack.mit.edu/ast/arrays/mwa}} \citep{bowman2013science, tingay2013murchison, dillon2014overcoming} and  Hydrogen Epoch of Reionization Array (HERA)\footnote{\url{http://reionization.org}} \citep{furlanetto2009cosmology, deboer2017hydrogen}. This signal, however, is predicted to be very weak and is buried in foregrounds that are $4$ to $5$ orders of magnitude larger than the expected signal (e.g. \citep{ali2008foregrounds, ghosh2012characterizing}). Despite sustained efforts, a detection is still forthcoming. The best upper limits on the mean-squared 21-cm brightness temperature fluctuations are $457~\text{mK}^2$ at $k=0.34~h\;\text{Mpc}^{-1}$ at $z=7.9$ and $3496~\text{mK}^2$ at $k=0.36~ h\;\text{Mpc}^{-1}$ at $z=10.4$, which come from HERA \citep{Abdurashidova_2023}.

% The redshifted  21-cm signal provides a three dimensional (3D) view of the Universe, where the distance along the  line-of-sight (LoS) direction is inferred from the measured frequency (or redshift).  However, this 3D view is not an instantaneous snapshot as the light-cone (LC) effect arising from  the finite speed of light implies that the look-back time increases with distance along the LoS  direction.  The mean \HI density  fraction $\bxHi$ evolves significantly with redshift during EoR, causing the  statistical properties of the fluctuations in the redshifted  21-cm signal to evolve with distance along the LoS direction.

The redshifted  21-cm signal provides a three dimensional (3D) view of the Universe, where the distance along the line-of-sight (LoS) direction is inferred from the measured frequency (or redshift). However, this 3D view is not an instantaneous snapshot, as the light-cone (LC) effect arising from the finite speed of light implies that the look-back time increases with distance along the LoS direction. This effect is particularly important during the EoR where $\bxHi$ the mean \HI neutral fraction decreases rapidly as reionization proceeds. Further, the sizes of ionized bubbles also increase as reionization proceeds. Both of these effects together cause the statistics of the LC EoR 21-cm signal to evolve with redshift, or equivalently,  
along the LoS direction. Considering the implications for observations to quantify the EoR 21-cm signal, there is a direct relationship between the bandwidth under consideration and the significance of the LC effect, and this becomes more significant with increasing bandwidth. \cite{datta2014light} and \cite{mondal2018} have shown that on large-scales the LC effect can lead to signal loss of up to 400\% in estimates of the EoR 21-cm power spectrum.

A pioneering work \cite{barkana2006light} considered a simple reionization model with spherical ionized bubbles whose characteristic radius $R_{\rm CH}$ and the corresponding $\bxHi$ both vary with redshift. They showed that the 21-cm two-point correlation function (2PCF) is anisotropic with respect to the LoS direction due to the LC effect. This signature of the LC effect was later verified  using large-scale numerical simulations by \cite{zawada2014light}. \cite{datta2012light, datta2014light} have carried out similar analyses on LC simulations using the 3D power spectrum (PS). The main issue addressed in all of these works is that $\bxHi$, and also the  statistical properties of the 21-cm fluctuations, evolve with distance along the LoS direction. They found that this affects the correlation function or the power spectrum at large length-scales. It is also important to note another LC effect that occurs even if $\bxHi$ and $R_{\rm CH}$ do not vary with $z$. The finite light travel time across an expanding bubble makes it appear elongated towards the observer along the LoS \citep{Sethi_2008}. This affects the 21-cm signal at small length-scales and  makes the power spectrum anisotropic along the LoS. Note that the present work is primarily concerned with the former ``large-scale" LC effect due to the evolution of $\bxHi$ and the statistical properties of the signal, and not the latter ``small-scale" LC effect due to the rapid bubble growth. 

It is important to note that the above-mentioned works that have used either the PS or the 2PCF all assume that the signal is statistically homogeneous (or ergodic) along the LoS direction. However, this assumption breaks down for the EoR LC 21-cm signal, for which $\bxHi$ evolves significantly along the LoS. As a consequence, both the PS and the 2PCF provide a biased estimate of the 2-point statistics of the EoR LC 21-cm signal \citep{trott2016exploring}. The Multi-frequency Angular Power Spectrum (MAPS) $\cl (\nu_1, \nu_2)$  \citep{datta2007multifrequency} overcomes this limitation because it does not assume that the signal to be ergodic along the LoS direction. \citep{mondal2018} have analyzed EoR LC 21-cm simulations to demonstrate that MAPS is able to successfully quantify the non-ergodic component of the signal, which is missed by the 3D PS $P(k)$. 

Redshift space distortion (RSD) is another important LoS effect that arises due to the peculiar velocities of \HI \citep{bharadwaj2004cosmic}. While substantial efforts have been invested in accurately including RSD in simulated  EoR 21-cm snapshots \citep{mao2012redshift, jensen2013probing, majumdar2013effect}, we still find it necessary to include a brief discussion here on how to incorporate them in LC simulations.  It is important to note that the LC effect, which depends on the look-back time, where the look-back time varies with LoS distance, depends on the actual distance  and not the redshift space distance, which has a peculiar velocity contribution. Therefore, it is necessary to first construct the light cone (LC) before applying the RSD to the \HI distribution, as detailed in \citep{mondal2018}.

To simulate the LC EoR 21-cm signal the total LoS  extent ($L$) is divided into a discrete number  ($\nrs$)  of intervals ($\Delta r_i$) each corresponding to a different look-back time  $t_i$ ($i=1,...,\nrs$). One proceeds by simulating the evolution of the \HI distribution in a cubic box  of dimension  $L$. This is used to produce snapshots of the \HI distribution, both position and velocity, corresponding to the time instances $t_i$.  We refer to each such coeval simulation as a reionization snapshot (RS), and $\nrs$ (introduced earlier)  denotes the number of reionization snapshots used for the LC simulation. Finally, slices from the reionization snapshots are stitched together along the LoS to produce the LC 21-cm signal. While \citep{mondal2018} have presented a method to accurately incorporate RSD and the LC effect in EoR  LC 21-cm simulations, 
neither they nor any of the previous works \cite{zawada2014light, rajat2009, datta2012light, datta2014light, Zhao_2022, mondal2019method, mondal2021} have provided any discussion or criteria for the number of snapshots $\nrs$ that need to be used, and the choice of $\nrs$ has been ad hoc. Considering the number of snapshots $\nrs$ used per redshift interval  in various earlier studies, this is $\le 10$ for  \cite{zawada2014light, datta2014light, rajat2009},  $\approx 13$ for \cite{datta2012light} and in the range $20 - 25$  for \cite{mondal2018,mondal2019method, mondal2021, Zhao_2022}. Many of these works have also interpolated the simulated brightness temperature maps to obtain values at intermediate redshifts between the  snapshots. While the typical source lifetimes of $\sim 10$ Myr (e.g. \cite{rajat2009}) or the ionized bubble growth time-scale of $\sim 8$ Myr \citep{majumdar2011} could be used to determine $\nrs$ and the time interval between snapshots, we note that these timescales do not necessarily characterize the evolution of $\bxHi$, which is the quantity  relevant for our considerations. In the present work, for the first time, we analyze and quantify the error in the LC 21-cm signal due to the choice of $\nrs$. We address the question ``What is the smallest $\nrs$ for which we can achieve the desired accuracy (say $5 \%$)  in modeling the LC effect?'' Along the way, we also present a new method for speeding up the LC simulations while retaining accuracy.

This paper is structured as follows: In Section~\ref{sec2} we briefly discuss the process of generating reionization snapshots. In Section~\ref{sec:LC_sim} we explain how these are stitched together to form a 21-cm LC simulation.  We present  the statistical tools used for our study in Section~\ref{sec4} and  we discuss the effect of a finite number of snapshots in the LC simulation in Section~\ref{effect}. We  introduce a new method to substantially  reduce the error without incurring much computational burden in section~\ref{mitigation}. We conclude and summarise our results in Section~\ref{sec6}.
This paper has used Plank+WP best-fitting values of the cosmological parameters $h=0.6704$, $\Omega_{\rm m0} = 0.3183$, $\Omega_{\Lambda 0} = 0.6817$, $\Omega_{\rm b 0} h^2 = 0.022 032$, $\sigma_8 = 0.8347$ and $n_{\rm s} = 0.9619$ \citep{collaboration2020planck}.

\section{Simulating the EoR 21-cm Signal}
\label{sec2}

We first discuss the process of generating the coeval reionization snapshots. This comprises {\bf three} major steps. In the {\bf first} step, we generate the dark matter density field at the desired redshifts using a particle mesh (PM) $N$-body code\footnote{\url{https://github.com/rajeshmondal18/N-body}} \citep{bharadwaj2004hi, mondal2015}. We have considered a comoving volume of [286.72 Mpc]$^3$ and the force is calculated using a 4096$^3$ mesh with 70 kpc grid spacing. The dark matter particle distribution is generated using 2048$^3$ particles which corresponds to a mass resolution of $1.09 \times 10^8 \,{\rm M}_{\odot}$. 
%\textbf{The time resolution of our N-body simulation is $\delta a =0.004$, where $a$ is the cosmological scale factor.} 
In the {\bf second} step, we identify collapsed
%the collapsed objects in the 
dark matter 
%particle distribution
halos using a Friends-of-Friend (FoF) halo finder\footnote{\url{https://github.com/rajeshmondal18/FoF-Halo-finder}} \citep{mondal2015, mondal2016}. The linking length for the FoF algorithm is set to be 0.2 times the mean inter-particle distance. A group of particles is considered a  halo if it consists of a minimum of 10 particles. Setting the minimum halo mass at only 10 particles is possibly not very reliable, and 20 particles are possibly a more conservative choice. However, as presented in Appendix \ref{app:b},  we find that our halo mass function obtained with 10 particles is in good agreement with the \cite{sheth2002excursion} mass function. This sets the minimum mass for a halo at $ 1.09 \times 10^9 \,{\rm M}_{\odot}$. The first two steps outlined above provide us with a set of ``cosmology snapshots (CS)'', one at each desired redshift. Each cosmology snapshot consists of the dark matter particle positions, peculiar velocities and a halo catalogue. In the subsequent discussion, we use $\ncs$ to denote the number of cosmology snapshots used to construct the LC EoR 21-cm simulations. Note that our fiducial value $\ncs=26$  corresponds to a time interval $\sim 2.01 \, {\rm Myr}$ which is smaller than the bubble growth timescale which is $\sim 8 \, {\rm Myr}$ at $z = 7.458$. 

\begin{figure*}
    \centering
    \includegraphics[width=\textwidth]{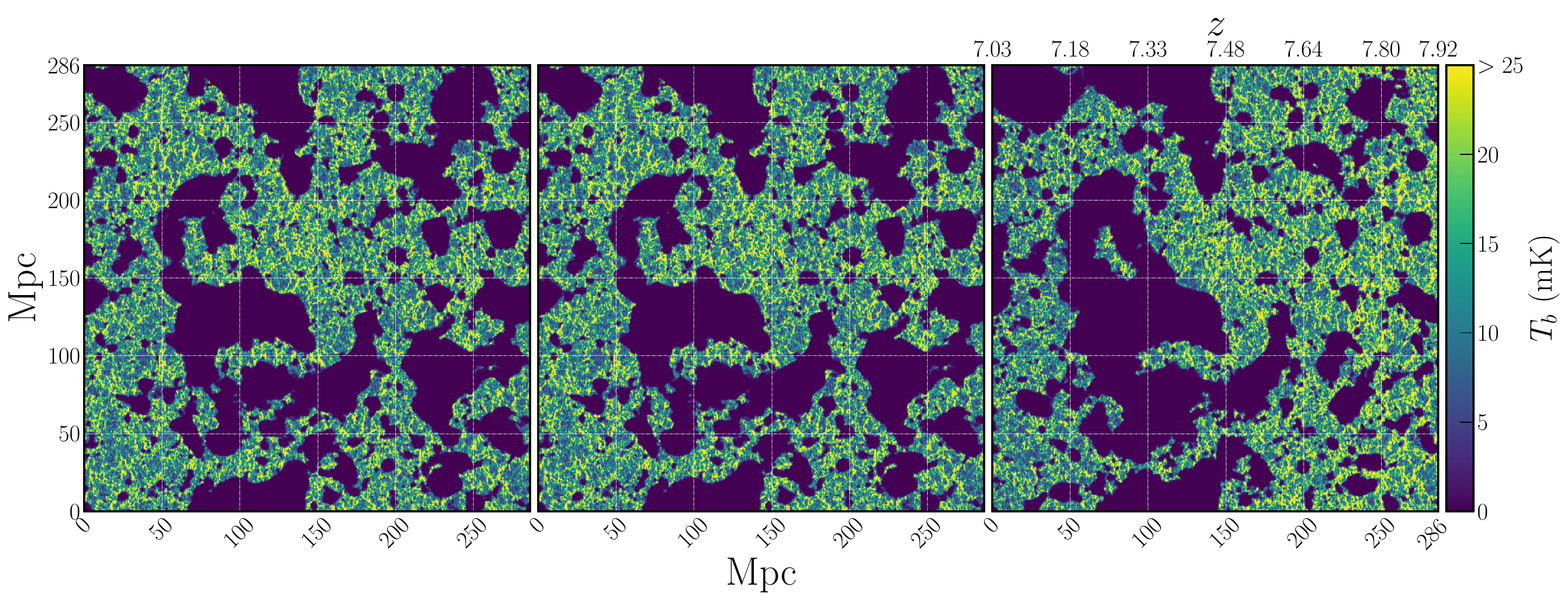}
    \caption{The brightness temperature maps $T_{ b}$ for the reionization snapshot without sampling (left), reionization snapshot with sampling (middle) and light-cone simulation using $\nrs = 26$ (right). All three simulations are centred at $z=7.458$.}
    \label{Tb}
\end{figure*}

In the {\bf third}  step, we generate the reionization snapshots (RS)  using a semi-numerical reionization code\footnote{\url{ https://github.com/rajeshmondal18/ReionYuga}} (see e.g. \citep{mondal2018, mondal2019method}). This closely follows the excursion-set formalism of \cite{furlanetto2004growth},  and employs the homogeneous recombination scheme as used in \cite{choudhury2009inside}. Our reionization model assumes that the hydrogen follows the underlying dark matter distribution, and the ionizing sources are hosted in the dark matter halos whose masses  exceed  a minimum halo mass $M_{\rm min}$. Further, the number of ionizing photons emitted by a source is assumed to be directly proportional to the mass of its host halo with a proportionality constant $N_{\rm ion}$. In addition to $M_{\rm min}$ and $N_{\rm ion}$, our reionization simulations have a third free parameter, namely the mean free path of the ionizing photon $R_{\rm mfp}$ (see e.g. \citep{mondal2018, Shaw2020} for a detailed description of the parameters). Setting different values of these parameters creates different reionization histories. For the present work, we have used the  parameter values  $N_{\rm ion} = 23.21$, $R_{\rm mfp} = 20$ Mpc and $M_{\rm min} = 1.09 \times 10^9 \,{\rm M}_{\odot}$, which results in a complete reionization by $z \sim 6$ and a 50\% ionization by $z \sim 8$ (see e.g. figure~1 of \citep{mondal2018}). This reionization history is consistent with \cite{davies2018quantitative}.
%and the corresponding integrated Thomson scattering optical depth $\tau = 0.057$ agrees with the observed CMB optical depth $\tau = 0.058 \pm 0.012$ \cite{collaboration2020planck}. 

Each RS  consists of a 3D \HI density field represented on a grid. Note that this grid is $8$ times coarser than that  for force calculation in the $N$-body simulations. We next account for the RSD, which  redistributes the \HI along the LoS direction according to 
\begin{equation} 
s=r+  \frac{v}{a H(a)}
\label{eq:rsd}
\end{equation}
where $r$ and $s$ are respectively the actual and redshift space distances to any particular \HI element, $v$ is its peculiar velocity along LoS and $a$ is the cosmological scale factor. In order to implement this, we represent
the simulated  \HI density field using \HI particles (see e.g. \citep{majumdar2013effect}) whose  positions and  peculiar velocities  are the same as the particles in the $N$-body simulation. The neutral hydrogen fractions $\xHI$\, from eight adjacent grid points are interpolated in each particle position to calculate their \HI masses. 
We then apply eq.~(\ref{eq:rsd}) to calculate the redshift space positions of the \HI particle, and then regrid them to calculate $\rho_{\rm \Hi}$, the \HI density in redshift space. Following \cite{majumdar2013effect} (see also \cite{bharadwaj2005using}), we calculate the \HI 21-cm brightness temperature distribution using 
\begin{equation}
   \Tb (\hat{n}, \nu) = \frac{\bar{T}_0}{\bar{\rho}_{{\rm H}}}  \rho_{\text{\Hi}} (\hat{{n}},s)
   \label{eq:tb1}
\end{equation}
and 
\begin{equation}
   \bar{T}_0 = 4.0 \:\text{mK} \:\left(\frac{\Omega_b h^2}{0.02}\right) \left(\frac{0.7}{h}\right)
\end{equation}
where $\bar{\rho}_{\rm H} $ is the mean hydrogen density. 
 
\begin{figure}
    \centering
    \includegraphics[width=0.48\textwidth]{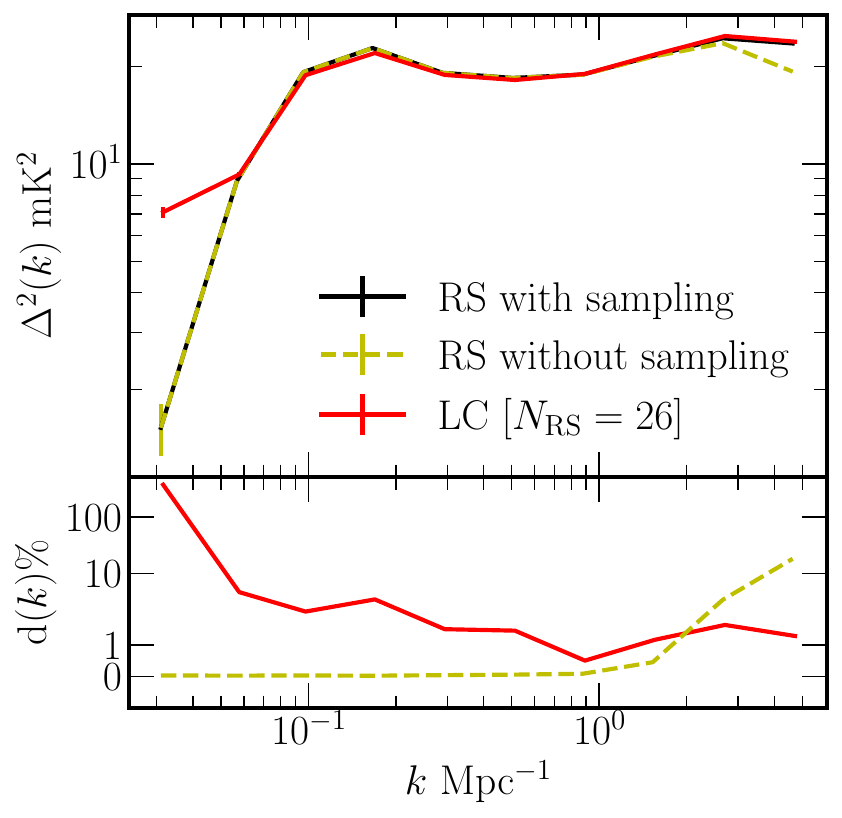}
    \caption{The top panel shows the spherically averaged 21-cm power spectrum $\Delta^2(k)$ and its 1$\sigma$ error bars for reionization snapshots with (black) and without (yellow) sampling, and for LC(26) simulation (red). The error bars are estimated assuming Gaussian statistics.  The red solid line in the bottom panel shows the percentage difference ${\rm d}(k) =\left(\left|\Delta^2_{\rm LC}-\Delta^2_{\rm RS}\right|/{\Delta^2_{\rm RS}}\right) \times 100\%$ between the LC and sampled RS, whereas, the yellow dashed line shows the same quantity for sampled and un-sampled RS.} 
    \label{pk_lc_cb}
\end{figure}

In order to reduce the computation, we have further randomly sampled the $N$-body particle distribution to bring down the number of \HI particles by a factor of eight. This substantially cuts down the computation time and memory requirement. The left and middle panels of 
Figure~\ref{Tb} show the 21-cm brightness temperature maps for a reionization snapshot at $z=7.458$ without and with the particle sampling, respectively. We notice that the two images are practically indistinguishable. The power spectra for these two images are shown in the top panel of Figure~\ref{pk_lc_cb}, using yellow dashed and black solid lines, respectively. The percentage difference between these two PS are shown in the bottom panel of the same figure, using yellow dashed line. We see that sampling affects the final power spectrum $P(k)$ only at the largest $k$ values $(k \sim 4 \, {\rm Mpc}^{-1})$ where there is a $\sim 20 \%$ increment, and the change is less than $1 \%$ for $k \le  2 \, {\rm Mpc}^{-1}$.

\subsection{Simulating the Lightcone 21-cm Signal}
\label{sec:LC_sim}

As mentioned earlier, it is necessary to first construct the LC and then incorporate the effect of the peculiar velocities. To do this, we start with the reionization snapshots  prior to applying the peculiar velocities. We then represent the \HI field in each snapshots using \HI particles, as discussed earlier and detailed in \citep{majumdar2013effect}.
The near ($n$) and far ($f$) ends of  the simulated  LC EoR 21-cm signal respectively correspond to the  frequencies (redshifts)  $\nu_n = 176.8$\,MHz  ($z_n = 7.03$) and $\nu_f = 159.2$\,MHz  ($z_f=7.92$) prior to incorporating  the effect of the peculiar velocities.   This  extends from $r_n = 8842.53$\,Mpc to $r_f = 9129.25$\,Mpc (comoving) along the LoS, which exactly matches the extent  of $N$-body simulation.  
Following \cite{mondal2018}, we have divided the LoS extent $r_f$ to $r_n$ into $\nrs$ intervals, each of comoving extent $\Delta r_i$ along the LoS. Each  interval  $\Delta r_i$  corresponds to a different look back  time $t_i$ or equivalently redshift $z_i$ with $i=1,2,3,...,\nrs$. We have prepared a different reionization snapshot for each $z_i$, so we have $\nrs$  reionization snapshots for  our LC simulation. Each interval $\Delta r_i$  of the LC is populated using the \HI particles drawn from  the corresponding reionization snapshot.

We calculate the observed frequency $\nu$ corresponding to the  $\nu_e=1420 \, {\rm MHz}$ emission from each \HI particle using  its  comoving distance  $r $ and LoS peculiar velocities $v$. Following \citep{mondal2018}, we have used  
\begin{equation}
    \Tb (\hat{{n}}, \nu) = \Bar{T_0} \frac{\rho_{ \text{\Hi}}}{\Bar{\rho}_{\text{H}}} \left(\frac{H_0 \nu_e}{c}\right) \left|\frac{\delta r}{\delta \nu}\right|
\end{equation}
to calculate the 21-cm brightness temperature $\Tb (\n, \nu)$ as a function of  the observed  frequency $\nu$ and the LoS direction $\n$.  The reader is referred to \citep{mondal2018} for details of the implementation of this step. 

We note that the inclusion of the peculiar velocity results in some of the \HI particles having frequency values beyond the boundaries of the LC box. This causes a lowering of the \HI particle density near the boundaries. 
We calculate the frequency interval that is affected by this particle reduction, and exclude slices of this size from both the nearest and farthest sides of the LC box. After accounting for the RSD, the final LC box extends from $\nu_f = 159.3$ MHz to $\nu_n= 176.6$ MHz and has a central frequency of $\nu_c = 167.94$ MHz, which corresponds to a redshift  $z_c = 7.458$; %however, we do not expect this to be a large effect.
Finally, we use the flat-sky approximation to  map the simulated 21-cm brightness temperature distribution to a 3D $(\thetavec, \nu)$  rectangular grid.  

The right panel of Figure~\ref{Tb} shows the brightness temperature distribution for the LC simulation using $\nrs=\ncs=26$ which is the fiducial value for the present analysis. Comparing this with the reionization snapshots (left and middle panels), we can identify the same ionized bubbles in all the maps. However, we see that the bubble sizes are different. Considering the LC simulation, the bubble sizes evolve along the LoS direction ($x$ direction) and they are smaller at high~$z$ and the sizes increase as reionization proceeds (low~$z$). This clearly illustrates the fact that the EoR LC 21-cm signal is not statistically homogeneous along the LoS direction.

\section{Statistical analysis}\label{sec4}

\begin{figure}
    \centering
    \includegraphics[width=0.48\textwidth]{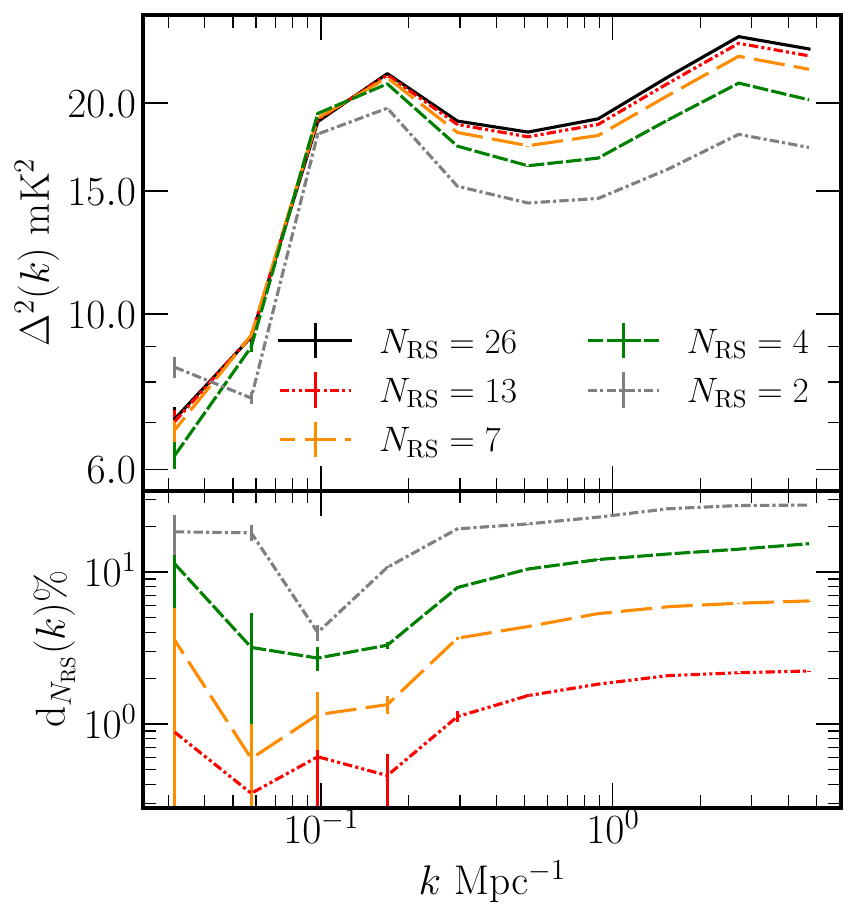}
    \caption{Top panel shows spherically averaged 21-cm 3D power spectrum for LC simulations using $\nrs =$ 26, 13, 7, 4 and 2. The bottom panel shows the percentage difference ${\rm d}_{\nrs}  (k)$ (eq.~\ref{eq:dncbk}).}
    \label{pk_lc_comparison}
\end{figure}

\subsection{The 3D power spectrum}
\label{sec:pk}
The spherically averaged 3D power spectrum $P(k)$ can be used to quantify the two-point statistics of the LC EoR 21-cm signal. The 21-cm brightness temperature distribution $\Tb (\n, \nu)$ can be expressed as a function of the observing frequency $\nu$ and the angle in the plane of the sky $\thetavec$. We map the $(\thetavec, \nu)$ coordinate to $(\bm{r}_\perp, r_\parallel)$ which are the comoving displacements  from the center of the simulation box respectively  perpendicular and parallel to  the LoS.  Here we have used $\bm{r}_\perp  = r_c \thetavec$ and  $r_\parallel = r'_c(\nu - \nu_c)$ where $r_c$ and $r'_c = \frac{d r}{d \nu}|_{r_c}$ are both  evaluated at $z_c$ which corresponds to $\nu_c$ the central frequency of the simulation. The result of this approximation is a spatially uniform rectangular grid where we can directly use 3D FFT to calculate $\tilde{T}_b(\bm{k})$ from $\Tb (\bm{r}_\perp, r_\parallel)$. This approximation costs less than 2\% error in grid positions. 

%From the $\tilde{T}_b$ the 3D \Hi 21-cm power spectrum can be estimated using
%\begin{equation}
%    P(\bm{k}) = V^{-1} \left<\tilde{T}_b(\bm{k}) \: \tilde{T}_b(\bm{-k})\right>.
%\end{equation}
The dimensionless spherically averaged 21-cm power spectrum $\Delta^2(k) = k^3 P(k) /2\pi^2$ is shown in Figure~\ref{pk_lc_cb} for the LC simulation and reionization snapshot both at central redshift $z_c$. We see that the LC effect significantly enhances $\Delta^2(k)$ at large scales. The LC and RS power spectra differ by $365\%$ at the smallest $k$ bin $k=0.03$ Mpc$^{-1}$. 
%However, at larger $k$ values the difference is less than 4\%. \textcolor{red}{In Figure~\ref{pk_lc_cb}, we also show the 1$\sigma$ error bars estimated using...}\textcolor{green}{the inverse of the square root of number of modes in a bin} Here we  note that the large-scale modes are affected by sample variance due to the finite simulation size. The actual value  might be somewhat different from the factor of $3.65$ quoted here, however, we note that this value is consistent with \cite{mondal2018}. 
% By definition, $P(k)$ estimates the two-point statistics averaged over all three directions. It is also important to note that LC 21-cm signal $ \Tb(\bm{\theta}, \nu)$ is not statistically homogeneous (ergodic) and evolves significantly along the line-of-sight direction (Figure~\ref{Tb}). 

% The Fourier transform also imposes periodicity on the signal which cannot be justified along the LoS in the presence of the LC effect. Due to this, the power spectrum fails to quantify the entire two-point statistics of the signal and gives a biased/incomplete estimate \citep{trott2016exploring, mondal2018}. 

\begin{figure*}
\centering
\includegraphics[width =\textwidth]{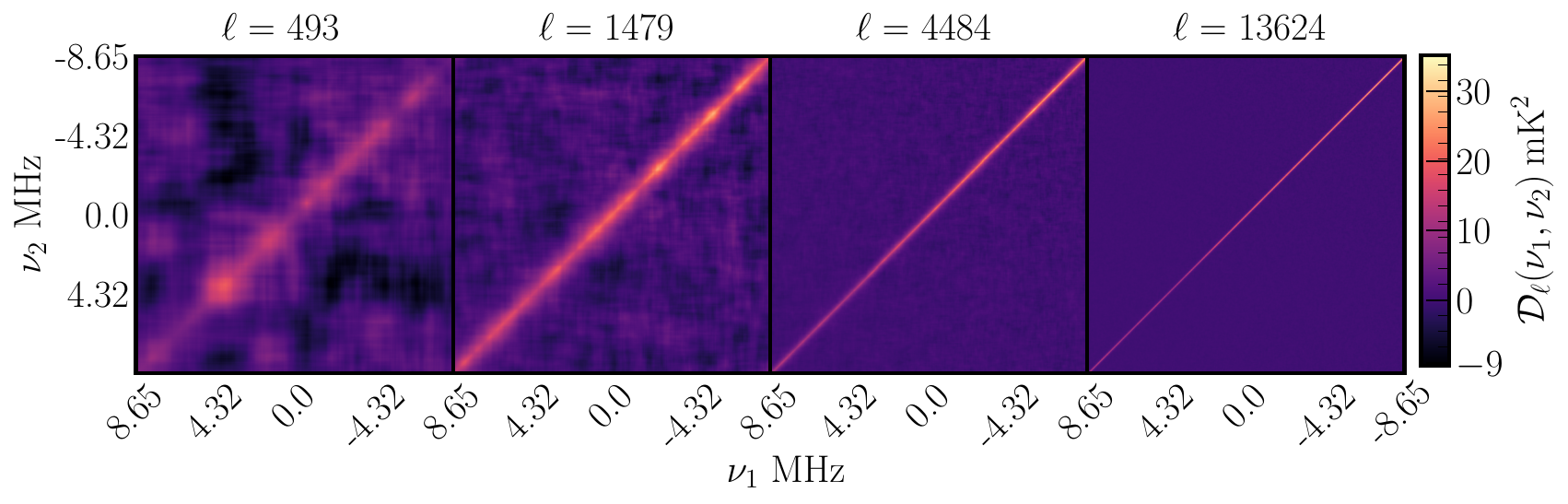}
\caption{This shows dimensionless MAPS $\dl(\nu_1,\nu_2)$ for the LC(26). MAPS for other LC($\nrs$) with $\nrs<26$ looks very similar, hence not shown here explicitly. }
\label{MAPS0}
\end{figure*}

\subsection{The Multi-frequency Angular Power Spectrum}
In contrast to the 3D power spectrum, the Multi-frequency Angular Power Spectrum (MAPS) $\cl (\nu_1, \nu_2)$ quantifies the entire second-order statistical information in the EoR 21-cm signal in the presence of LC effect (see e.g. \citep{mondal2018, Mondal2022}). 
%We decompose the LC 21-cm brightness temperature fluctuations $\delta T_b (\bm{\hat{n}}, \nu)$ in the basis of spherical harmonics $Y_l^m(\bm{\hat{n}})$ as
%\begin{equation}
%    \delta T_b (\bm{\hat{n}}) = \sum_{\ell, m} a_{\ell m } (\nu) Y_{\ell}^m(\bm{\hat{n}})
%\end{equation}
%and define the MAPS \citep{datta2007multifrequency} as
%\begin{equation}
%    \cl (\nu_1, \nu_2) = \left<a_{\ell m} (\nu_1) a^{*}_{\ell m}(\nu_2)\right>
%\end{equation}
%In this work we adopt the flat-sky approximation. Under this assumption 
The $\bm{\theta}$ dependence of $\Tb (\bm{\theta}, \nu)$ can be decomposed into 2D Fourier modes. The 2D Fourier transform of $\Tb (\bm{\theta}, \nu)$ is $\tTb (\bm{U}, \nu)$. Subsequently, the MAPS is defined as
\begin{equation}
    \cl (\nu_1,\nu_2) = {\mathcal C}_{\rm 2 \pi U} (\nu_1, \nu_2) = \frac{1}{\Omega}\left<\tTb (\bm{U}, \nu_1) \tTb(-\bm{U}, \nu_2) \right>
\end{equation}
where, $\bm{U}$ is the Fourier conjugate of $\bm{\theta}$ with  $|\bm{U}| = \ell / 2\pi$. Here, $\Omega$ is the solid angle subtended by the simulation sky at the observer point. The only assumption here is that the EoR 21-cm signal is statistically homogeneous and isotropic along different directions on the sky plane. However, no such intrinsic assumption is made along the LoS direction $\nu$.  

The $\ell$ range corresponds to our LC simulations span from $\ell_{\rm min} = 2\pi /\theta_{\rm max} = 200$ to $\ell_{\rm max} = 2\pi/\theta_{\rm min} = 51215$. We divided this range into 10 equally spaced logarithmic bins. We calculated the average $\cl (\nu_1, \nu_2)$ in each of these bins. For our analysis, we have chosen four $\ell$ bins, $\ell = 493, 1479, 4484 \;\text{and}\; 13624$ to show our results. These $\ell$ bins correspond to large scale, two intermediate scales and small scale, respectively. %We also use the dimensionless MAPS $\dl(\nu_1, \nu_2) = \ell (\ell + 1) \cl (\nu_1, \nu_2)/2\pi$ to show our results.

\begin{figure*}
\centering
\includegraphics[width =\textwidth]{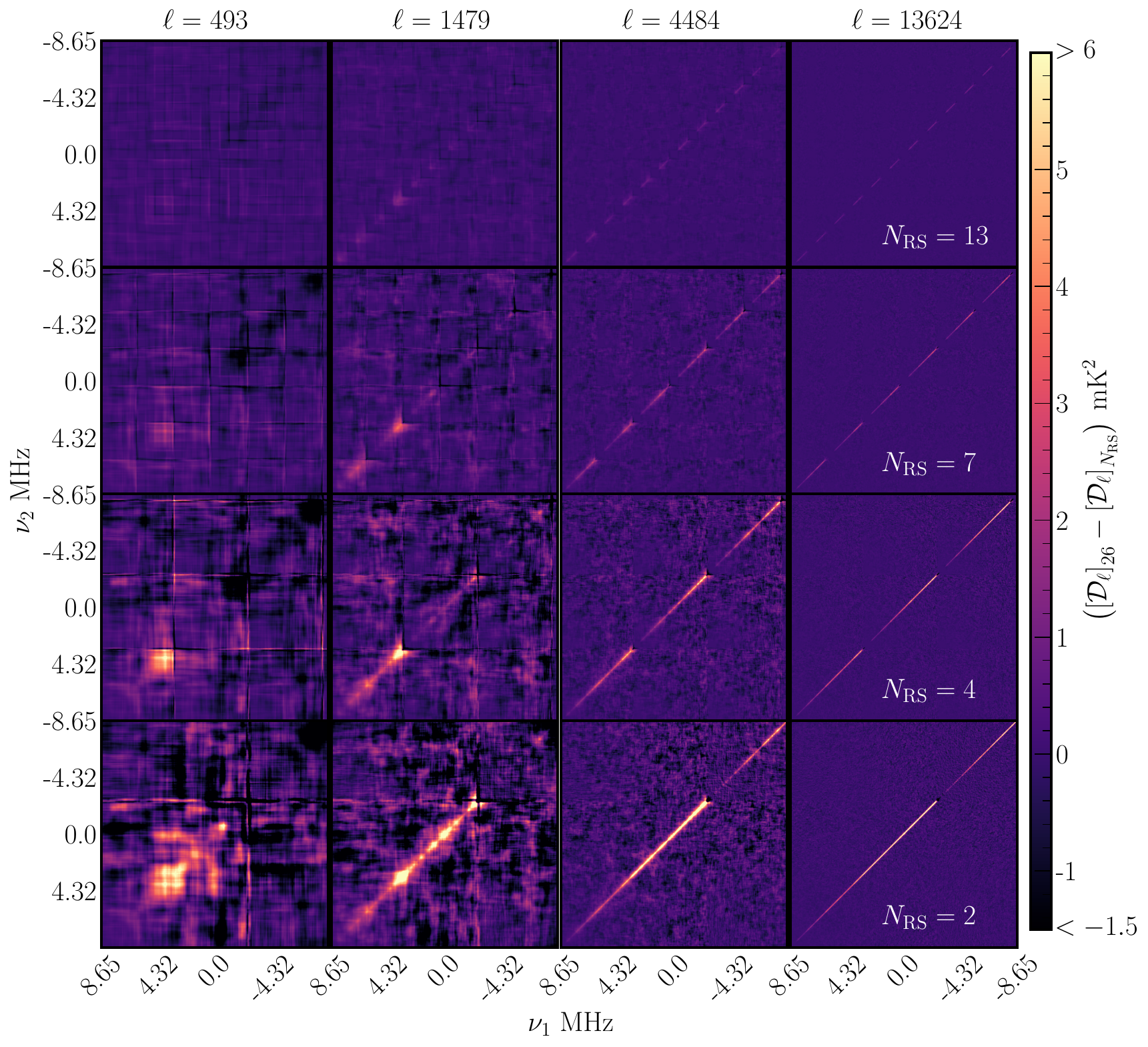}
\caption{The difference in dimensionless MAPS $\dl(\nu_1, \nu_2)$ for LC simulations using different $\nrs<26$ with respect to LC(26) simulation: $\left[\dl\right]_{26} - \left[\dl\right]_{\nrs}$.}
\label{MAPS}
\end{figure*}

% \begin{figure*}
% \centering
% \includegraphics[width =\textwidth]{plots/Diag.pdf}
% \caption{This shows the diagonal elements of MAPS $\dl (\nu, \nu) {\rm mK}^2$ for lightcone simulations using different $\ncb$ and at different $\ell$ values.}
% \label{Diag}
% \end{figure*}

\begin{figure*}
\centering
\includegraphics[width =\textwidth]{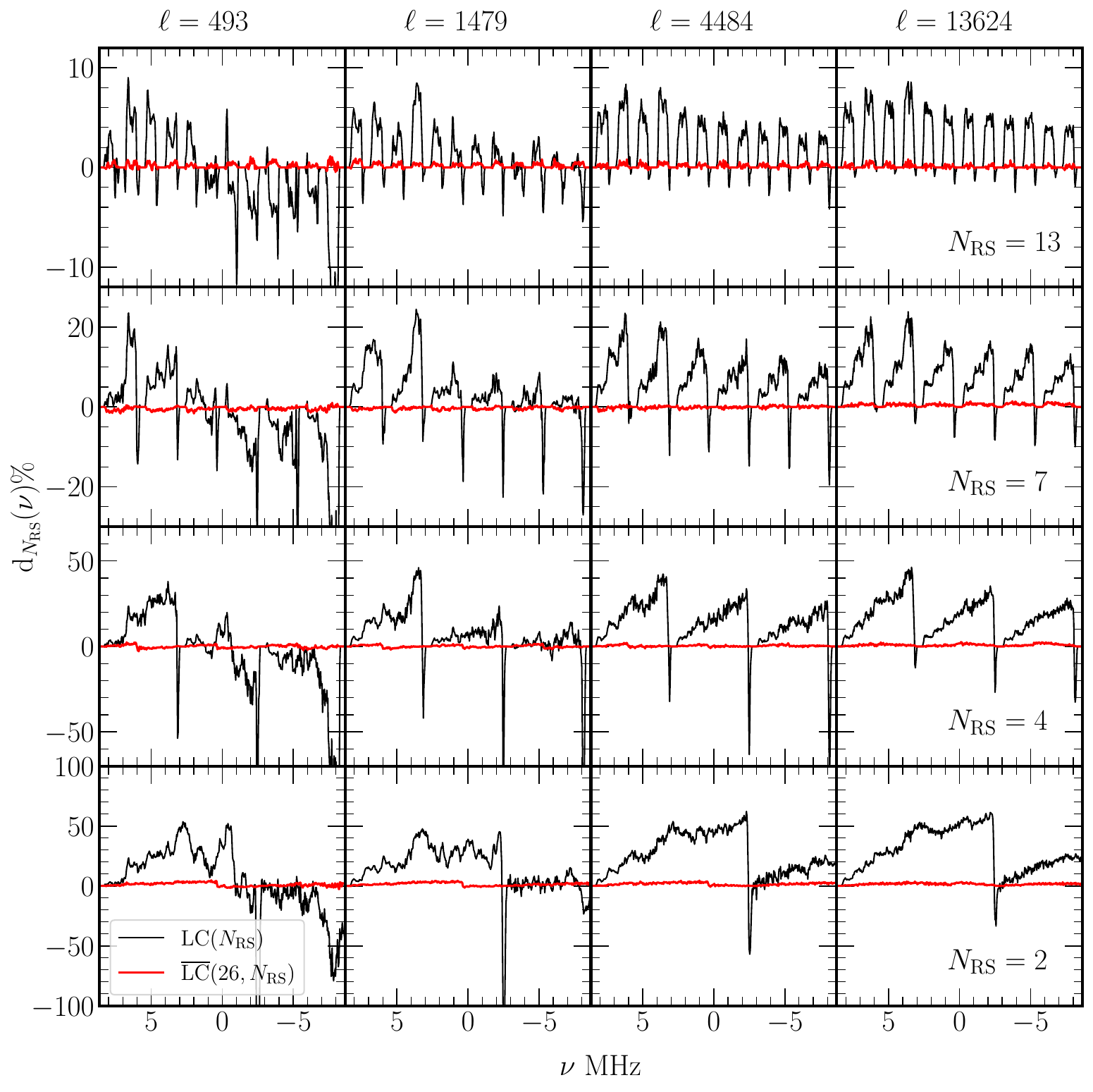}
\caption{The percentage difference of the diagonal elements of MAPS $\dl(\nu, \nu)$ for different LC simulations with respect to LC(26) ${\rm d}_{\nrs} (\nu)$ (eq.~\ref{eq:dncbnu}). The results are shown both for before (black) and after (red) mitigation.}
\label{Diag_diff_all}
\end{figure*}

%\section{Results}
\section{Effects of snapshot intervals}
\label{effect}
We have constructed the LC simulation by stitching together $\nrs$ slices, each 
from a different  reionization snapshot, using the methodology presented in Section~\ref{sec:LC_sim}. The value of $\bxHi$ changes abruptly at the stitching boundaries, and 
 $\delta \bxHi$, the jumps in $\bxHi$, decreases as $\nrs$, the number of reionization snapshots, is increased. Here we find $\delta \bxHi \approx 0.2, 0.1, 0.05$ and $0.02$ for $\nrs = 2, 4, 7$ and $13$, respectively. The discontinuity is within $1 \%$  $(\delta \bxHi \le 0.01)$ for  $\nrs = 26$,  and we use this as the reference to compare other LC simulations. Our interest here is to quantify the errors in the statistics of the LC EoR 21-cm signal due the finite value of $\nrs$.  For this, we have considered the spherically averaged 3D power spectrum $P(k)$ and the MAPS $\cl (\nu_1, \nu_2)$. We have considered LC simulation with $\nrs =$ 26, 13, 7, 4 and 2, and we refer to these as LC$(\nrs)$ {\it i.e.} LC(26), ..., LC(2), respectively. Note that the number of cosmology snapshots ($\ncs$) is exactly equal to the number of reionization snapshots ($\nrs$) for all the LC simulations discussed till now, and it is not necessary to explicitly refer to $\ncs$ in the current discussion. 
 Note that the LC simulations are all of the same size and span the same $z$ range mentioned in Section~\ref{sec:LC_sim}. 

We first discuss the results for the 3D power spectrum. The upper panel of Figure~\ref{pk_lc_comparison} shows the dimensionless 21-cm power spectrum $\Delta^2(k)$ for LC simulations with different values of $\nrs$. Here we spherically bin the $k$ range of our simulation into $10$ equally spaced logarithmic bins. The effects of using different $\nrs$ are significant at both the large and the small scales. We see that for all values of $k$  the value of $\Delta^2(k)$ decreases as  $\nrs$ is reduced. Conversely, the power spectrum approaches its ``true value'' as  $\nrs$ is increased.  We also notice that the values of $\Delta^2(k)$ appear to  have converged by $\nrs=26$ as  we find only a small increment in $\Delta^2(k)$  as $\nrs$ is increased from $13$ to $26$. The error bars corresponding to cosmic variance are shown in the figure. The bottom panel of Figure~\ref{pk_lc_comparison} shows the percentage difference in  $\Delta^2 (k)$ for different LC$(\nrs)$  relative to  LC(26) 
\begin{equation}
    {\rm d}_{\nrs} (k)  = \frac{\left|\Delta^2_{26} (k)-\Delta^2_{\nrs} (k)\right|}{ {\Delta^2_{26} (k)}} \times 100\%.
    \label{eq:dncbk}
\end{equation}
The overall value of ${\rm d}_{\nrs} (k)$ at all $k$ bins increases as we decrease $\nrs$. The value of ${\rm d}_{2}$ is $\sim18\%$ at small $k$ bins and $\sim 27.6\%$ at the largest $k$ bin, whereas the value of ${\rm d}_{13}$ is $<1\%$ for small $k$ bins and $\sim 2.2\%$ at the largest $k$ bin. 
%The $\delta^{26}_{\ncb}$ for LC(13) is almost zero at $k = 0.09\,{\rm Mpc}^{-1}$ which reaches $0.18$ at $k = 2.70\,{\rm Mpc}^{-1}$. For LC(2) the errors at these scales are $0.1$ and $0.27$, respectively. 
The average of ${\rm d}_{\nrs}$ for all $k$ values are $19\%$, $9\%$, $3\%$ and $1\%$ for LC(2), LC(4), LC(7) and LC(13) respectively. Although the value of the 3D power spectrum changes with $\nrs$, it does not provide any insight into what exactly causes this difference. This is because  $P(k)$ is spherically averaged over all directions,  whereas the use of different $\nrs$ introduces discontinuities only along the LoS direction as we will explore next.

Figure~\ref{MAPS0} shows the dimensionless MAPS $\dl (\nu_1, \nu_2 ) = \ell(\ell+1)\cl(\nu_1, \nu_2)/2\pi$ for the LC(26) simulation. For the convenience of plotting, we have shifted the origin of the frequency scale to $\nu_c$ (i.e. $\nu \rightarrow \nu -\nu_c$)  throughout. As noted in several earlier works (e.g. \citep{datta2007multifrequency, mondal2019method}), we see that the signal is largely localized in  the vicinity of the diagonal elements $\nu_1 = \nu_2$. The signal falls off rapidly away from the diagonal, and it has a very small value  $\dl (\nu_1, \nu_2 ) \approx 0$ for large frequency separations $\Delta \nu =\mid \nu_1 - \nu_2 \mid$. We also notice that this fall-off with increasing $\Delta \nu$ occurs faster for larger $\ell$ (e.g. \citep{Mondal2020}). For $\ell = 493$ the value of $\dl$ falls close  to zero for $\Delta \nu \sim 4$\,MHz, whereas this is $\Delta \nu \sim 0.25$\,MHz for $\ell=13624$. The figures showing $\dl (\nu_1 , \nu_2 )$ for smaller values of $\nrs$ are very similar, and it is very difficult to visually  distinguish these from $\nrs=26$  and  we have not explicitly shown them here. However, it is possible to visualize these differences by considering $\left([\dl(\nu_1, \nu_2)]_{26} - [\dl(\nu_1, \nu_2)]_{\nrs}\right)$ shown in Figure~\ref{MAPS}. We see that the differences are visible  for all the $\ell$ values, and these increase as  $\nrs$ is reduced. Further, the differences appear to follow a systematic pattern mostly close to the diagonal $\nu_1 = \nu_2$, where the MAPS signal peaks. These are particularly prominent close to  the ``stitching boundaries'' which correspond to the locations where we have joined two adjacent reionization snapshots.

To further illustrate the discontinuities  due to the limited $\nrs$, we focus on the diagonal elements of MAPS {\it i.e.} $\dl (\nu, \nu)$. In Figure~\ref{Diag_diff_all}, we show  
\begin{equation}
    {\rm d}_{\nrs}(\nu)  = \frac{\left[\dl (\nu, \nu)\right]_{26} - \left[\dl (\nu, \nu)\right]_{\nrs}}{\left[\dl (\nu, \nu)\right]_{26}} \times 100\% \, ,
    \label{eq:dncbnu}
\end{equation}
which quantifies the percentage difference of $\dl (\nu,\nu)$ for LC($\nrs$) relative to LC(26). Considering the black lines in Figure~\ref{Diag_diff_all}, as expected we find $\nrs -1$ number of sharp discontinuities in ${\rm d}_{\nrs}(\nu)$, one  corresponding to each  stitching boundary between two adjacent reionization snapshots. The peak values of  ${\rm d}_{\nrs}(\nu)$  are listed  in Table~\ref{tab:error} for the $\nrs$  and $\ell$ values for which the results are shown in Figure~\ref{Diag_diff_all}.  Considering $\nrs=2$, we find  ${\rm d}_{\nrs}(\nu)  \approx  -241 \%$ at the stitching boundary for  the smallest $\ell$ bin. The magnitude of the discontinuity goes down, but it is still quite large ($\mid {\rm d}_{\nrs}(\nu) \mid > 50 \%$) for the larger $\ell$ bins. Considering $\nrs=4$, we see that the magnitude of the discontinuities is comparable to those for $\nrs=2$. However, considering the mean $\mid {\rm d}_{\nrs}(\nu) \mid$ (shown within $(...)$ in Table~\ref{tab:error}), we see that this decreases when $\nrs$ is increased from $2$ to $4$. For example, the mean $\mid {\rm d}_{\nrs}(\nu) \mid$ drops from $22.8 \%$ to $16.1 \%$ at the smallest $\ell$ bins when $\nrs$ is increased from $2$ to $4$. Considering $\nrs=7$ and $13$, we see that the peak ${\rm d}_{\nrs}(\nu)$ and the   mean $\mid {\rm d}_{\nrs}(\nu) \mid$ both go down as $\nrs$ is increased. We find a peak ${\rm d}_{\nrs}(\nu) \approx -17 \%$ and mean  $\mid {\rm d}_{\nrs}(\nu) \mid \approx 2.4 \%$ at the smallest $\ell$ bin for $\nrs=13$. The peak ${\rm d}_{\nrs}(\nu)$ and the mean $\mid {\rm d}_{\nrs}(\nu) \mid$ have values of $ \sim 8 \% $ and  $\sim 2 \%$ respectively at the larger $\ell$ bins for $\nrs=13$.

We recollect that the aim here is to quantify the errors due to the finite number of reionization snapshots $\nrs$ in the LC simulation. Ideally, we should compare the results for a finite LC($\nrs$) with those from a continuum LC, unfortunately, the latter is not available.  Here we have used LC(26) as a proxy for the continuum LC, and we use  ${\rm d}_{\nrs}(\nu)$ to quantify the error due to a limited number of $\nrs$. We see that the error due to a finite $\nrs$ is mainly seen as a sharp discontinuity at the stitching boundary between two adjacent reionization snapshots. The amplitude 
(peak ${\rm d}_{\nrs}(\nu)$) of this discontinuity can be rather large, and it is typically found to be an order of magnitude larger than the  average  error which we have quantified using the mean  $\mid {\rm d}_{\nrs}(\nu) \mid $ (Table~\ref{tab:error}). It is interesting to note that the errors in the spherical power spectrum $\Delta ^2(k)$ (Figure~\ref{pk_lc_comparison}), which by definition is averaged over all directions, are comparable to the average errors in $\dl (\nu, \nu)$. The fact that the  spherical power spectrum is oblivious to the discontinuities at the stitching boundaries 
reiterates its limitations  to fully capture the statistics of the LC simulations.

\section{Mitigating the error}
\label{mitigation}
As discussed in Section~\ref{effect}, the error in the process of generating the LC 21-cm signal can be large due to the limited number  $\nrs$ of reionization snapshots. Here we propose a computationally economical method to mitigate this error by generating a large number of reionization  snapshots  at closely spaced time intervals.  The computationally expensive cosmology snapshots, however, are only generated at large time intervals and interpolated assuming a linear cosmological evolution.

%%%%%%%%%%%%%%%%%%%%%%%%%%%%%%%%%%%%%%%%%%%%%%%%%%%%%%%%%%%%%%%%%%%%%%%%%%%%%%%%%%%%%%%%%%%%%%%%%%%

\begin{table*}
\centering
\caption{Considering  LC(26) as the reference, this lists  the 
maximum (mean absolute) percentage difference (eq.~\ref{eq:dncbnu})
of $\dl (\nu,\nu)$ 
for A. LC($\nrs$) with  $\nrs=\ncs$ ; and B. $\overline{\rm LC}(26,\ncs)$.}
\label{tab:error}
\begin{tabular}{llllll}

\hline\vspace{-.2cm} \\

{} & {$\nrs \mid \ell$} & {493} & 1479 & 4484 & 13624 \\[4pt]

\hline\vspace{-.2cm} \\

{} & 13 & -17.1 (2.39) & 8.44 (1.56) & 8.34 (2.23) & 8.60 (2.56) \\[2pt]
{A.} & 7 & -65.9 (6.80) & -27.1 (4.73) & 23.5 (6.78) & 23.8 (7.44)\\[2pt]
{} & 4 & -268 (16.2) & -117 (10.2) & -76.2 (15.4) & 46.3 (16.7)\\[2pt]
{} & 2 & -241 (22.8) & -121 (18.1) & 62.3 (26.5) & 61.3 (28.8) \\[4pt]
\hline\vspace{-.2cm} \\

{} & 13 & 1.12 (0.18) & 0.89 (0.19) & 0.80 (0.15) & 0.82 (0.11) \\[2pt]
{} & 7 & -1.65 (0.46) & -1.42 (0.33) & -1.24 (0.21) & 1.43 (0.47)\\[2pt]
{B.} & 4 & -2.87 (0.66) & -1.87 (0.52) & 2.10 (0.40) & 2.57 (0.81)\\[2pt]
{} & 2 & 4.74 (1.45) & 4.32 (1.56) & 4.51 (1.61) & 3.76 (1.56) \\[4pt]
\hline
\end{tabular}
\end{table*}

%%%%%%%%%%%%%%%%%%%%%%%%%%%%%%%%%%%%%%%%%%%%%%%%%%%%%%%%%%%%%%%%%%%%%%%%%%%%%%%%%%%%%%%%%%%%%%%%%%%

We consider a situation where we have $\ncs$  cosmology snapshots {\it i.e.} coeval boxes after the first two steps of the simulation process described in Section~\ref{sec2}. As mentioned earlier, we have the dark matter particle positions, peculiar velocities and halo catalogue for each of the $\ncs$ cosmology snapshots which correspond to redshifts  $z_i$ with $i=1,2 ..., \ncs$. Note that this is the most computationally expensive step in our simulation process, and it is advantageous to keep the value of $\ncs$ as small as possible. The straight-forward way ahead, as discussed earlier, is to implement the third step of Section~\ref{sec2} and generate $\nrs$ reionization snapshots, one for each cosmology snapshot ($\nrs=\ncs$). We finally stitch these together to generate LC($\nrs$) using the methodology presented in Section~\ref{sec:LC_sim}. However, as already seen, we have large discontinuities (errors) in the LC simulation for small values of $\nrs$ (Section~\ref{effect}). 

\begin{figure*}
\centering
\includegraphics[width =0.8\textwidth]{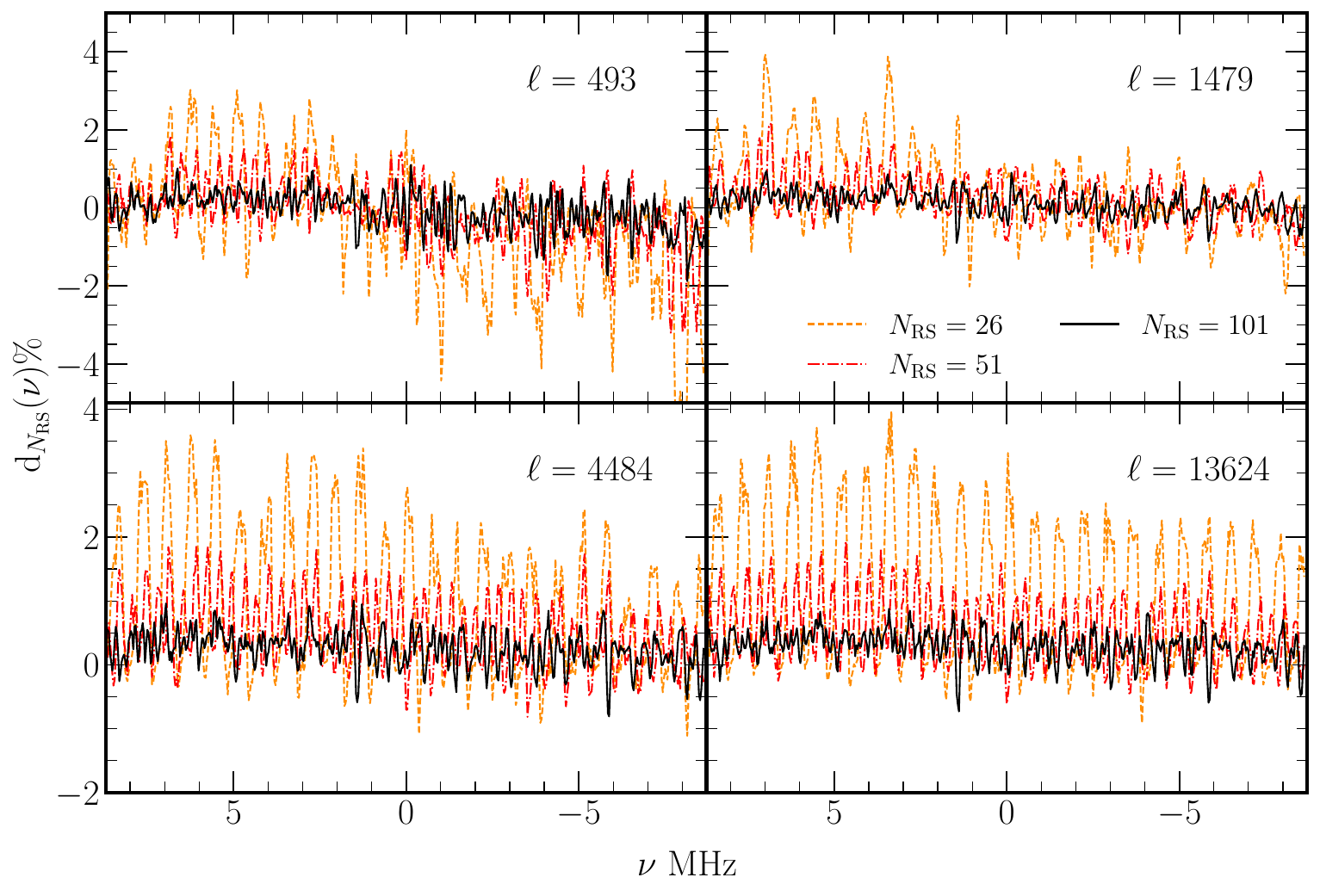}
\caption{The  percentage error in $\dl (\nu,\nu)$ for LC simulations with  $\nrs \ge 26$. Here ${\rm d}_{\nrs} (\nu)$ (eq. \ref{eq:dncbnu}) is the percentage error of LC($\nrs$) relative  to the LC simulation with the next higher value of $\nrs$.}
\label{Upper}
\end{figure*}

We now present a new procedure which also starts from the same $\ncs$ cosmology snapshots at $z_i$ with $i=1,2 ..., \ncs$, as mentioned in the previous paragraph. However, instead of straightaway proceeding to step 3 of Section~\ref{sec2}, we first grid the dark matter density and the halo mass density fields, and linearly interpolate these (in redshift) at $m$ redshift values between every successive pair $z_i$ and $z_{i+1}$.  This  gives us the dark matter density (also the H density) and halo mass density fields at a total of  $\ncs^{'}=[\ncs + m \times (\ncs-1)]$ intermediate redshifts  in place of the original $\ncs$. We now implement step 3 of Section~\ref{sec2} to generate $\nrs=\ncs^{'}$ reionization  snapshots, which we stitch together to produce a 21-cm LC simulation. We refer to this new LC simulation as the ``interpolated LC simulation" and denote this using $\overline{\rm LC}(\nrs,\ncs)$. The excursion set formalism which is used to generate the reionization  snapshots, and stitching the adjacent 21-cm slices are not computationally expensive steps. It is computationally inexpensive  to implement a large value of $m$ (and $\nrs$), provided $\ncs$ is small.  Note that we also need the dark matter particle positions and peculiar velocities at the $m$ intermediate $z$ values, here we have used the values from $z_i$. However, we have checked that using  $z_{i+1}$ also gives similar results.

We first validate the new procedure, and test how well the interpolated LC simulations $\overline{\rm LC}(\nrs,\ncs)$ match the original LC simulation LC($\nrs$). As earlier, we consider LC(26) as the reference and compare $\overline{\rm LC}(26,13)$ against the reference. Both  LC(26) and  $\overline{\rm LC}(26,13)$ have been constructed from 26 reionization snapshots. However, they differ because for LC(26) every reionization  snapshot also corresponds to a different cosmology snapshot, whereas we have used only $\ncs=13$ cosmology snapshots for $\overline{\rm LC}(26,13)$ where the other 13 intermediate reionization snapshots were generated by interpolating the H and halo density fields. We have similarly also generated $\overline{\rm LC}(26,\ncs)$ for $\ncs=2,4$ and $7$. The red lines in Figure~\ref{Diag_diff_all} show the percentage error (eq.~\ref{eq:dncbnu}) in $\dl(\nu,\nu)$ for   $\overline{\rm LC}(26,\ncs)$  with respect to  LC(26). We see that the errors have been considerably reduced,  and these are now very close to zero. The peak error and the mean absolute error are both listed in Table~\ref{tab:error}. The improvement achieved through the interpolated simulations is quite evident for all values of $\ncs$.  The magnitude of the  peak  errors is $<2 \%$ for $\ncs=13,7$ while this is $<3 \%$ for $\ncs=4$. Considering the extreme case where $\ncs=2$, we see that the magnitude of the peak error is $<5 \%$ for the interpolated LC simulations as against $\sim 200 \%$ for LC(2) where we do not interpolate. Note that the computational requirement for  $\overline{\rm LC}(26,2)$ is approximately sixteen times smaller than that for LC$(26)$, whereas the differences are less than $5 \%$ at maximum and $< 2 \%$ on average. It is quite evident that interpolated LC simulations are very close to the LC simulations provided we have the same number of reionization  snapshots.  The interpolation, however, allows us to significantly cut down the computational costs, and this provides a very economical and efficient technique to generate LC simulations.

We now use the new procedure described above to generate interpolated LC simulations using all the $\ncs=26$ available cosmology snapshots. We have considered $m=1,3$ and $7$ which provides us with interpolated LC simulations $\overline{\rm LC}(\nrs,26)$ with $\nrs=51,101$ and $201$ respectively. We expect these to  match the corresponding LC simulations to within a few percent (possibly better than $1 \%$) accuracies, and for the subsequent discussion, we drop the distinction and refer to these three interpolated LC simulations as   LC(51), LC(101) and LC(201) respectively. We now effectively have LC($\nrs)$  with $\nrs=[2,4,7,13,26,51,101,201]$, and we use ${\rm d}_{\nrs}(\nu)$ to quantify the percentage difference in $\dl(\nu,\nu)$ between any two  LC simulations  with successive $\nrs$ values. For example,  ${\rm d}_{26}(\nu)$ is the percentage difference of  $\nrs=26$ relative to $\nrs=51$. Considering  ${\rm d}_{26}(\nu)$ shown in Figure~\ref{Upper}, we see that we still have the discontinuities at the stitching boundaries, but the amplitude has come down considerably ($ < 4 \%$  everywhere barring one point where it is $\sim -7 \%$) as compared to the much larger values seen in Figure~\ref{Diag_diff_all}. Considering  even larger values of $\nrs$, we see that the amplitude of the discontinuities drops  further to  $\sim 2\%$ and $1 \%$ for ${\rm d}_{51}(\nu) $ and ${\rm d}_{101}(\nu)$ respectively.

We next consider  $\langle \mid {\rm d}_{\nrs} \mid \rangle$ which  is the mean absolute  ${\rm d}_{\nrs}(\nu)$ (averaged over $\nu$) for a fixed $\ell$ and $\nrs$.  This  provides an overall quantitative picture of the errors in the LC simulation due to the discrete number of reionization snapshots. Figure \ref{power_law} shows  $\langle \mid {\rm d}_{\nrs} \mid \rangle$ as a function of $\nrs$ for different values of $\ell$. For all values of $\ell$, we find the approximate scaling $\langle \mid {\rm d}_{\nrs} \mid \rangle \propto \nrs^{-1}$. Note that for the smallest $\ell$ bin, the scaling holds only if we discard the two smallest $\nrs$ values.  The  average absolute percentage error in $\Delta^2 (k)$ (Figure~\ref{pkvscb_k}) also shows a similar $\approx \nrs^{-1}$ scaling for the three largest $k$ bins, however, the smaller $k$ bins show a  steeper scaling (in the range $-2.5$ to $-1.5$).

\section{Summary and Conclusions.} 
\label{sec6}
During EoR, the mean neutral hydrogen fraction $\bxHi$ evolves significantly along the LoS direction due to the LC effect. It is essential to properly incorporate this in simulations of the EoR 21-cm signal. The LC 21-cm signal is typically simulated by stitching together a finite number ($\nrs$) of reionization snapshots, each corresponding to a different epoch. In this paper, we have quantified the errors  due to the finite value of $\nrs$. The simulations here span $286.7$ covering Mpc, which correspond to the redshift range $z=7.917$ to $7.032$ where $\bxHi$ evolves from $0.477$ to $0.157$. We use the diagonal elements of the dimensionless MAPS $\dl(\nu,\nu)$ to quantify the error. We find that the main effect is to introduce discontinuities in the simulated 21-cm signal at the stitching boundaries between two adjacent snapshots. Considering the extreme case with  $\nrs=2$ we find  a  peak error of $-241 \%$ at the stitching boundary where  $\bxHi$ jumps by $0.2$.  In this case, the mean absolute  error is $\sim 23 \%$, which is approximately an order of magnitude smaller.  Considering  LC$(13)$ we see that  the maximum jump in $\bxHi$ at the stitching boundaries is $0.02$, which is an order of magnitude lower compared to $\nrs=2$. In this case, the  peak error drops to  $-17.1 \% $ and the mean absolute error is  $\sim  3 \%$.  We observe that the error decreases as we increase $\nrs$ whereby abrupt changes in $\bxHi$ at the stitching boundaries become more gradual. However, the difficulty arises in that it  is computationally expensive to increase the number of snapshots that go into the 21-cm LC simulation. 

\begin{figure}
\centering
\includegraphics[width =0.48\textwidth]{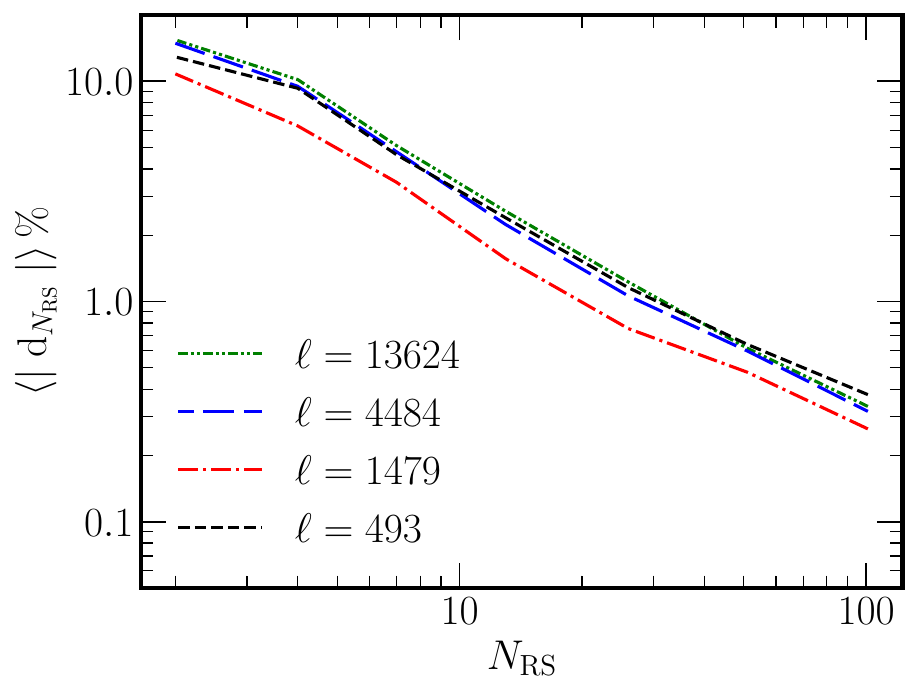}
\caption{The  mean absolute percentage error $\left<\mid{\rm d}_{\nrs}\mid\right>$ in $\dl (\nu,\nu)$ for LC$(\nrs)$ as a function of $\nrs$ for different $\ell$ values. }
\label{power_law}
\end{figure}

\begin{figure}
    \centering
    \includegraphics[width=0.48\textwidth]{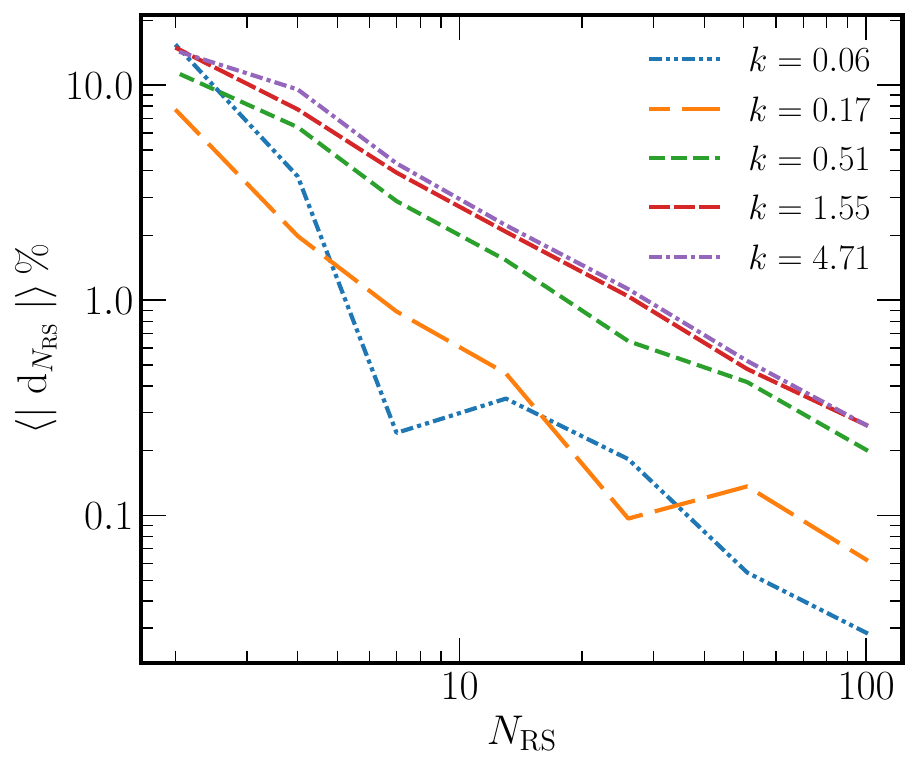}
    \caption{The mean absolute percentage error  in $\Delta^2(k)$ versus $\nrs$ for different $k$ bins.}
    \label{pkvscb_k}
\end{figure}

Analyzing the computational costs, the major expense goes into the first two steps (Section~\ref{sec2}) where we simulate  the cosmology snapshot which  jointly refers to the dark matter particle positions, peculiar velocities and halo catalogue. It is relatively inexpensive to apply  the excursion set formalism (the third step in Section~\ref{sec2}) on a cosmology snapshot to generate  the  reionization snapshot. Typically we use $\ncs =\nrs$  where $\ncs$ is the number of cosmology snapshots {\it i.e.} we generate a separate  cosmology snapshot  for  each reionization snapshot.   The limitation  for increasing the accuracy of 21-cm LC simulations comes from the computationally expensive cosmology snapshots which limit the value of $\nrs$.

In this paper we introduce a technique to increase $\nrs$ keeping $\ncs$ fixed. The underlying idea is to generate $\ncs$ cosmology snapshots at a few  well separated redshifts, and then interpolate the gridded dark matter and halo fields to several ($\nrs$) closely spaced intermediate redshifts for which we simulate reionization snapshots. We have demonstrated this technique by interpolating  only two cosmology snapshots ({\it ie.} $\ncs=2$) to produce 26 reionization snapshots ($\nrs=26$). Comparing the resulting LC simulation with another which has $\ncs=\nrs=26$, we find that the two match to better than  $5 \%$ accuracy (Figure~\ref{Diag_diff_all} and Table~\ref{tab:error}). Considering  $\ncs=4$ further increases the accuracy to within $3 \%$. This clearly establishes our interpolation scheme as a cost-effective and highly accurate technique to increase the number of reionization snapshots for LC simulations. Proceeding further, we have fixed $\ncs=26$ and used the interpolation scheme to construct LC simulations with $\nrs=51, 101$ and $201$. Comparing the results for $\nrs=101$  with those for $201$ (Figure \ref{Upper}), we see that the relative differences are less than $\sim 1 \%$ which indicates a high level of convergence for the LC simulation. We expect a very small improvement ($ < 1 \% $) if $\nrs$ is increased any further. We have also studied how the mean absolute error $\left<\mid{\rm d}_{\nrs} \mid\right>$ for the LC simulation changes if $\nrs$ is varied. We find that this shows  a scaling $\left<\mid{\rm d}_{\nrs}\mid\right> \propto \nrs^{-1}$ (Figure~\ref{power_law}) for all the $\ell$ bins considered here. Similar behaviour is also observed at large $k$ for the 3D spherical power spectrum (Figure~\ref{pkvscb_k}).

Considering earlier works which have studied the LC effect, we note that several of these \cite{rajat2009,zawada2014light, datta2014light} have respectively used $\nrs \approx 8, 10, 3$ (per unit $z$ interval). The corresponding time intervals are approximately $6, 5, 17$ Myr which are comparable to and sometimes larger than the bubble growth timescale of $\sim 8$ Myr. Note that  the peak error is predicted to be $\sim 70 \%$ for $\nrs=7$. However, these works, and several other works \citep{datta2012light,Zhao_2022}, have interpolated  the brightness temperature distribution along the LoS direction to generate values for intermediate redshifts  where  snapshots are not available. While this will smooth the discontinuities at the stitching boundaries, it is not guaranteed that result will approach the actual \HI distribution. This is because the ionization fronts which partition the volume into two distinct regions, one  neutral and the other ionized. These fronts propagate into the neutral region as reionization progresses. Interpolating the brightness temperature distribution will not capture the propagation of these fronts, and will make them appear thicker instead. In contrast, our method interpolates  the hydrogen and ionization source distribution, and calculates the ionization state for every reionization snapshots. As demonstrated here,  this  provides a computationally inexpensive technique to accurately model the \HI distribution.

In conclusion,  it is important  to  accurately incorporate the LC effect  in simulations of the EoR 21-cm signal.  This is particularly relevant in order to correctly interpret the signal when it is detected.  Here we have quantified the errors in the 21-cm LC simulations due to the finite number of snapshots. We have shown how these errors can be controlled by increasing the number of snapshots, and we have also demonstrated a technique to achieve this within  limited computational expenses. 
%{\bf Several} earlier predictions of the EoR 21-cm LC signal have used a  limited number of snapshots.      

It is necessary to construct large ensembles of EoR 21-cm  simulations in order to make reliable predictions for the statistics  of   various quantities like the 3D power spectrum, bispectrum  or MAPS  (e.g. \citep{mondal2015,mondal2016}). However, currently  the predictions for EoR LC 21-cm simulations are largely limited to a few  statistically independent realizations.  
In future work we propose to use multiple realizations of EoR LC 21-cm simulations to analyze the statistics of the expected 21-cm signal and  make predictions for ongoing and future EoR experiments. We also  plan to use the technique developed here to perform accurate 21-cm LC simulations for a variety of reionization models. 
 It may be noted that the present work has only considered  a particular source model. The results will change  if the source model is varied, we however expect the main conclusions of this work to hold irrespective of source model.

\appendix
\section{Errors in 3D power spectrum}
Figure~\ref{pkvscb_k} shows $\langle \mid  {\rm d}_{\nrs} \mid \rangle$ the mean absolute  percentage error in $\Delta^2(k)$ for 21 cm LC simulations with different $\nrs$. We find that this scales as $\langle \mid  {\rm d}_{\nrs} \mid \rangle \propto \nrs^{-1}$ for  the three larger $k$ bins. However, the scaling is steeper with  exponent $-2.31$ and $-1.84$ for the smallest and the second smallest $k$ bins respectively. Notice that this is different from the behaviour in Figure~\ref{power_law} where we have a $-1$ scaling for all the $\ell$ bins.

\section{The halo mass function}
\label{app:b}

\begin{figure}
\centering
\includegraphics[width=0.48\textwidth]{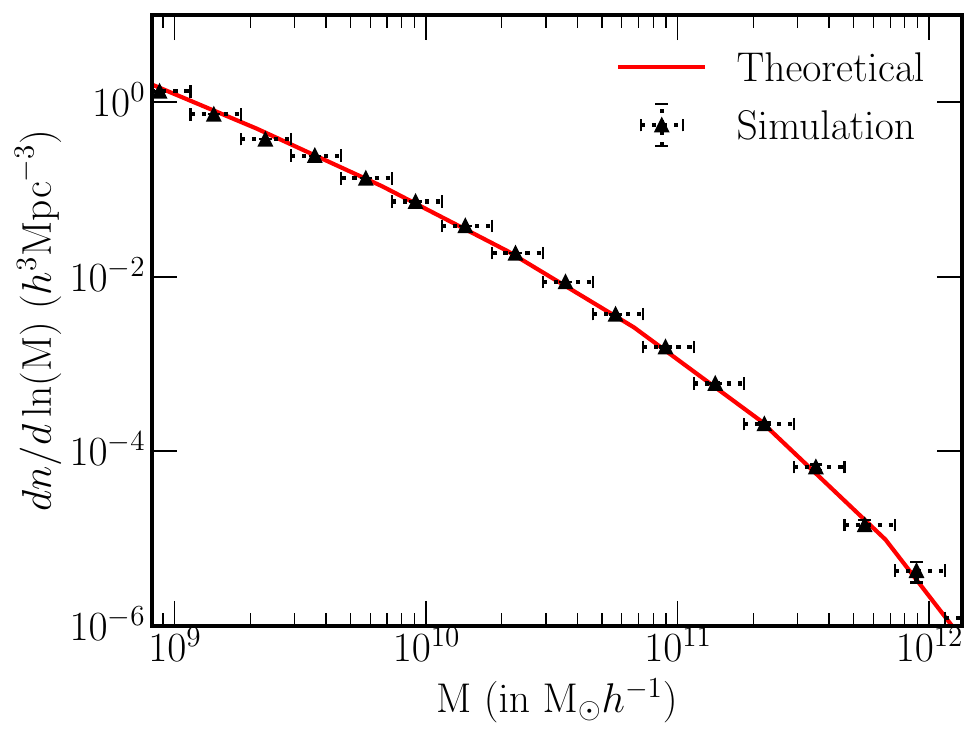}
\caption{This shows the theoretical \citep{sheth2002excursion} and simulated halo mass function
at redshift $z=7.458$.}
\label{mat_halo}
\end{figure}

%The left-hand panel of 
Figure~\ref{mat_halo} shows the theoretical \citep{sheth2002excursion} and simulated halo mass function at $z=7.458$. The halos were identified using a minimum of 10 dark matter particles which corresponds to a 
minimum halo mass of $ 1.09 \times 10^9 \,{\rm M}_{\odot}$.  We  note that resolving halos with 10 particles is not ideal, particularly for a PM $N$-body code, where a group of $\sim 50$ particles is generally used to reliably identify a halo. However, in many simulations using the same code, \cite{mondal2015, mondal2016, mondal2018, mondal2019method, Shaw2019, Shaw2020} have found that the halo mass function obtained using a minimum group of 10 particles for halo identification is in good agreement with theoretical prediction, as shown in Figure~\ref{mat_halo}.

% \acknowledgments
\section*{Acknowledgements}
SP acknowledges support from the Prime Minister's Research Fellowship (PMRF). SP would like to thank Abinash Kumar Shaw and Kh. Md. Asif Elahi for their help. SP and SB acknowledge the super-computing facilities at the Centre for Theoretical Studies, Department of Physics, IIT Kharagpur.

% \begin{theunbibliography}{}
\bibliography{LC.bib}

% \end{theunbibliography}

\end{document}